\documentclass[unnumsec,webpdf,contemporary,large,authoryear]{oup-authoring-template}%

\theoremstyle{thmstyleone}%
\theoremstyle{thmstyletwo}%
\theoremstyle{thmstylethree}%

\usepackage{dsfont}
\usepackage{algorithm}
\usepackage{algpseudocode}
\usepackage{amsmath} 
\usepackage{mathtools}
\begin{document}
	
	\journaltitle{Bioinformatics}
	\DOI{doi.}
	\copyrightyear{2022}
	\pubyear{2025}
	\access{Advance Access Publication Date: Day Month Year}
	\appnotes{Paper}
	
	\firstpage{1}
	

\title[A review on the longitudinal omics data analysis]{Longitudinal Omics Data Analysis: A Review on Models, Algorithms, and Tools}

\author[1]{Ali R. Taheriyoun\ORCID{0000-0002-2282-7741}}
\author[1]{Allen Ross\ORCID{0009-0004-4777-2419}}
\author[2]{Abolfazl Safikhani\ORCID{0000-0001-8678-1247}}
\author[3]{Damoon Soudbakhsh\ORCID{0000-0002-9313-8804}}
\author[1,$\ast$]{Ali Rahnavard\ORCID{0000-0002-9710-0248}}

\authormark{Taheriyoun et al.}

\address[1]{\orgdiv{Department of Biostatistics and Bioinformatics}, \orgname{the George Washington University}, \orgaddress{\street{Washington}, \postcode{20052}, \state{DC}, \country{USA}}}
\address[2]{\orgdiv{Department of Statistics}, \orgname{George Mason University}, \orgaddress{\street{Fairfax}, \postcode{22030}, \state{VA}, \country{USA}}}
\address[3]{\orgdiv{Department of Mechanical Engineering}, \orgname{Temple University}, \orgaddress{\street{Philadelphia}, \postcode{19122}, \state{PA}, \country{USA}}}

\corresp[$\ast$]{Corresponding author. \href{email:rahnavard@gwu.edu}{rahnavard@gwu.edu}}




\abstract{Longitudinal omics data (LOD) analysis is essential for understanding the dynamics of biological processes and disease progression over time. This review explores various statistical and computational approaches for analyzing such data, emphasizing their applications and limitations. The main characteristics of longitudinal data, such as imbalancedness, high-dimensionality, and non-Gaussianity are discussed for modeling and hypothesis testing. We discuss the properties of linear mixed models (LMM) and generalized linear mixed models (GLMM) as foundation stones in LOD analyses and highlight their extensions to handle the obstacles in the frequentist and Bayesian frameworks. We differentiate in dynamic data analysis between time-course and longitudinal analyses, covering functional data analysis (FDA) and replication constraints. We explore classification techniques, single-cell as exemplary omics longitudinal studies, survival modeling, and multivariate methods for clinical/biomarker-based applications. Emerging topics, including data integration, clustering, and network-based modeling, are also discussed. We categorized the state-of-the-art approaches applicable to omics data, highlighting how they address the data features. This review serves as a guideline for researchers seeking robust strategies to analyze longitudinal omics data effectively, which is usually complex.}


\maketitle

\section{Introduction}
\begin{figure*}[!t]
	\includegraphics[width=.97\textwidth]{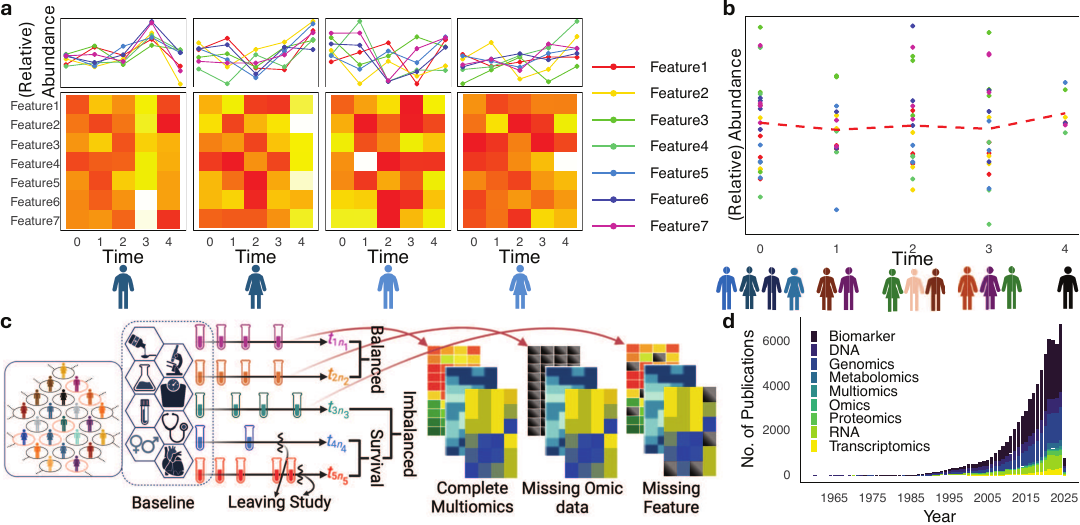}
	\caption{\textbf{Overall study designs and omics measurements. a}, shows the basic structure of longitudinal data with repeated measurements per subject. In contrast, \textbf{b}, shows a time-course study where `time' is a covariate without repeated measurements. \textbf{c}, illustrates multiomics studies with imbalanced sampling designs and missing data. The first two subjects are balanced despite irregular time-points, while subjects 4 and 5 drop out, leading to imbalance. \textbf{d}, shows the trend of publications based on longitudinal omics studies within the last 8 decades in PubMed databases generated by \href{https://github.com/omicsEye/pubSight}{\texttt{pubSight}}.}\label{overview}
\end{figure*}

Longitudinal data consist of repeated measurements from multiple subjects over time. Unlike time-course (time-series) data, which track a realization of a stochastic process, longitudinal data are sparse and subject-dependent (\autoref{overview}\textbf{a} and \textbf{b}). Biomarker interactions can enhance detection power, such as correlated metabolites in diabetes studies \citep{GuaschFerretal2016}. However, full multivariate models for serial measurements introduce high-dimensional estimation challenges. A practical alternative for univariate outcomes is incorporating random effects into fixed-effect models, such as the linear mixed model (LMM) \citep{LairdWare}.

Despite challenges, longitudinal studies provide robust statistical tests, reducing noise compared to independent samples \citep{ZegerLiang1992}, with lower variation in the estimates \citep{Liquetetal2012} and provide the progress path in treatment regimen. These have increased the demands for longitudinal omics data (LOD) in public health and pharmaceutical studies (\autoref{overview}\textbf{d}), exponentially. 

This review examines analysis methods for longitudinal omics studies, highlighting challenges in addressing key scientific questions. We summarize research questions, available datasets, methodologies, and computational tools. Studies tackling multiple issues are categorized and discussed accordingly, with emphasis on their application to LOD. Table S1 categorizes studies for current methods and applications, and \autoref{figsummaries} provides a summary of novel approaches and applications.
As highlighted in \autoref{overview}\textbf{c} and described bellow, LOD analyses present challenges beyond the classic LMM approaches:\\
\noindent\textbf{Imbalanced or missing:} In genomics, imbalances arise from missing data, mostly irregular time-points, or feature redundancy. Multiomics studies further complicate imbalances, as different omics layers may be missed. While most of the classic tools ignore the rows with missing values some; e.g., \texttt{JointAI} \citep{Erleretal2021} and \texttt{bild} \citep{Goncalvesetal2012} offer alternatives, though model specification remains critical.

\noindent\textbf{Correlated outcomes:} Omics studies exhibit intra-subject and inter-feature correlations. While standard univariate models ignore the latter, multivariate methods and random effects can account for these correlations.

\noindent\textbf{Time-varying covariates:} Clinical and demographic factors change over time, influencing variance and effect estimation. Ignoring such changes can distort statistical inference.

\noindent\textbf{Nonlinearity:} Biological systems are rarely linear, and while linear models are commonly used, they may fail to capture complex biological feedback.

\noindent\textbf{Leave-out:} Subject dropout in clinical studies, due to treatment changes, study completion, or mortality, introduces biases, especially if dependent on omics measurements.

\begin{figure*}[t]
	\centering
	\includegraphics[width=\linewidth]{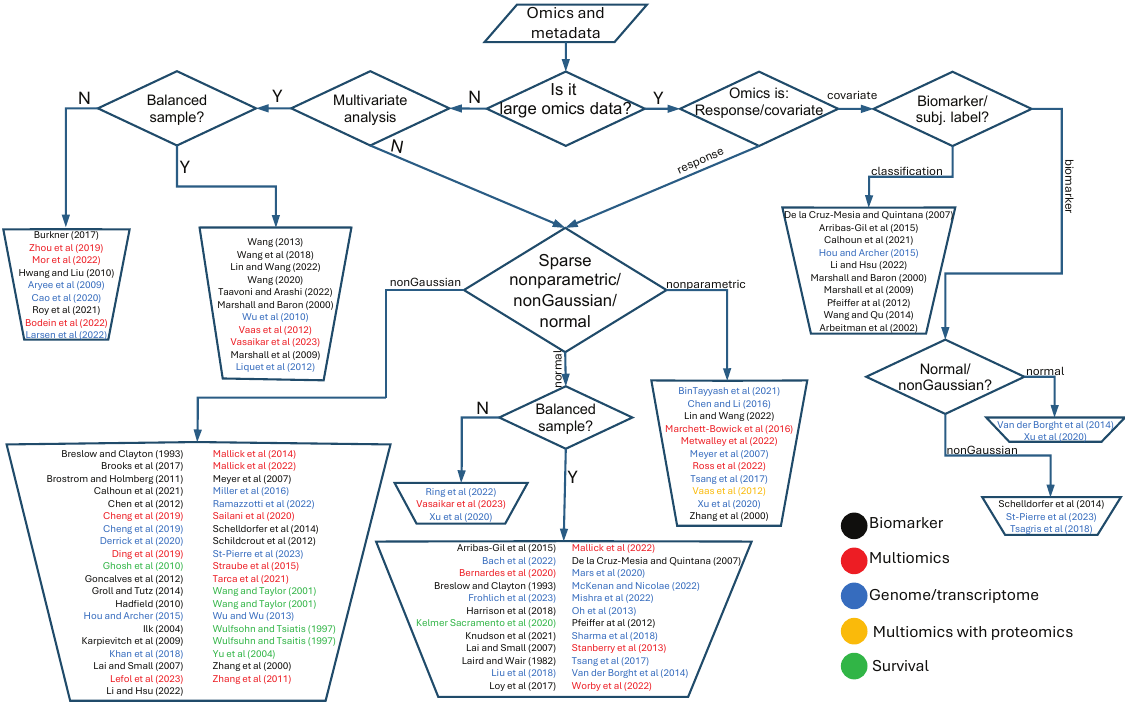}
	\caption{\textbf{General step-by-step workflow in a LOD analysis based on a supervised method.} It considers the multivariate methods for not large omics data, including the biomarker analysis or targeted metabolomics to achieve higher power. It also differentiates between the studies in that omics are studying under the effect of metadata (omics are response), and those omics are causing metadata changes (omics are covariates).}
	\label{guideline}
\end{figure*}

Addressing these challenges requires robust statistical methods, computational advancements, and refined study designs, which we discuss in subsequent sections. A brief list of reviewed studies implemented supervised methods in the analysis of LOD are categorized into a step-by-step flowchart in \autoref{guideline} (see also Algorithms \ref{alg;overall} and \ref{alg;FDR}). 

\section{LMM-based approaches for longitudinal analysis of omics}\label{continuous}
We can formulate an omics study that leverages linear mixed models methodologies. For an omics feature, let $\mathbf{y}_i$ represent the measurements (e.g., count, abundance, intensity) for the $i^\text{th}$ subject at time-points $t_{i1},\ldots,t_{in_i}$, with a corresponding design matrix $\mathbf{X}_i$ for fixed effects. Subject-specific random effects, $\mathbf{b}_i$, account for intra-subject correlations via a known design matrix, $\mathbf{Z}_i$ (see Section \ref{AppenLMMGLMM}):  
\begin{eqnarray}\label{LMM}
	\mathbf{y}_i=\mathbf{X}_i\boldsymbol{\beta}
	+\mathbf{Z}_i\mathbf{b}_i+ \boldsymbol{\varepsilon}_i,
\end{eqnarray}
where $\boldsymbol{\varepsilon}_i\overset{iid}{\sim} N(\mathbf{0},\boldsymbol{\Sigma}_{\mathrm{Err}})$ is Gaussian noise, independent of $\mathbf{b}_i \overset{iid}{\sim}N(\mathbf{0},\boldsymbol{\Sigma}_{\mathrm{rndEff}})$. Model estimation via restricted maximum likelihood is computationally accessible via for instance  \texttt{lme4} \citep{lme4}, \texttt{nlme} \citep{nlme}, and \texttt{proc mixed} in \texttt{SAS}. Ignoring random effects can bias RNA analyses \citep{Cuietal2016}.

LMM variants are widely used in repeated measures analysis and are crucial for studying longitudinal omics data \citep{Harrisonetal2018}. They help identify omics features that vary over time and across subjects while accounting for correlation. However, special considerations are needed for the type of distribution, missing data, time-varying covariates, or imbalanced data, which we review here.

\subsection{Linear mixed models: omics features under the normality assumptions}\label{LMM-based methods}

Alongside exploratory analysis \citep{RajeswaranBlackstone2024}, continuous omics features are commonly modeled using LMM, as seen in bulk and single-cell transcriptomics and proteomics studies \citep{Zhouetal2019,Tarcaetal2021,Tarcaetal2022,Bach2022,Frohlichetal2023}.
LMM partitions metadata into fixed effects (systematic influences) and random effects (within-subject correlation, e.g., sampling region or patient). Under normality assumptions, model \eqref{LMM} captures these variations, with residuals aiding exploratory analysis. For example, plotting residuals from an LMM (with time as a random effect) against those from an explanatory variable (e.g., a clinical variable) reveals clearer relationships.

LMM assumptions include independent subjects, independent random effects, and normality. However, diagnostic tests \citep{Loyetal2017} are often overlooked. Post-fitting, the type II Wald Chi-squared test \citep{Stroup2012} assesses abundance differences over time. Predictions at unobserved levels require diagnostics or uncertainty quantification even via bootstrapping.

In a study on LINE-1’s role in Parkinson’s, 423 patients with up to five follow-ups were analyzed \citep{Frohlichetal2023}. A clinical and transcriptomic dataset (114 response variables) was used, modeling subject effects as random effects. Each response was analyzed via LMM, with p-values adjusted using the Benjamini-Hochberg method. A similar approach was applied to longitudinal gut microbiome data \citep{Worbyetal2022}.

For greater flexibility, spline bases can replace squared time terms in LMM \citep{Straubeetal2015}. This involves filtering low-variance genes, regressing each feature on sampling time using spline LMM—where $f(t_{ij})$ replaces $\beta_0 + \beta_1t_{ij}$—and performing clustering and differential expression analysis based on the best-fit model.

\begin{figure}[h]
	\centering
	\includegraphics[width=.8\linewidth]{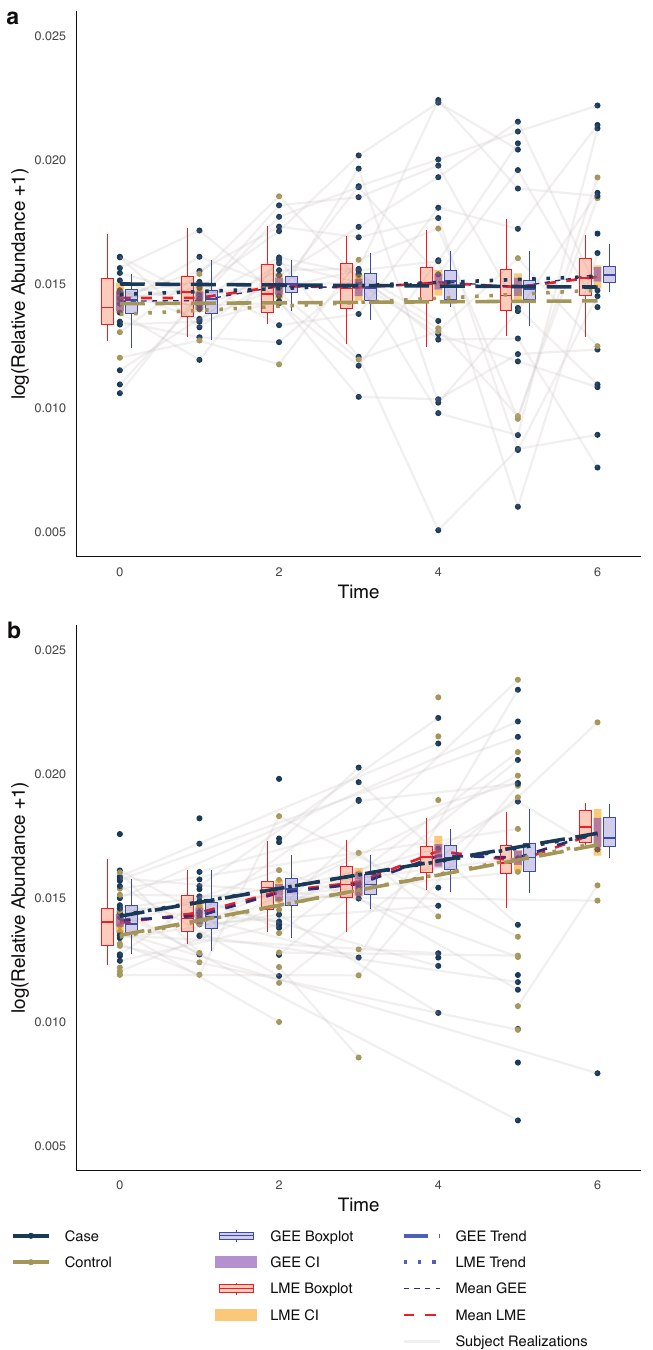}
	\caption{\textbf{The performance of LMM and GEE} is compared for \textbf{a}, balanced data from 20 subjects observed at seven time-points, and \textbf{b} imbalanced data of the same size but with varying observation times per subject. Treatment groups are distinguished by khaki (case) and navy blue (control). Mean fitted value curves are shown as dashed blue (GEE) and red (LMM) lines. The dynamics of $\log(\text{relative abundance}+1)$ over time are represented by dashed (GEE) and dotted (LMM) lines for each group. Boxplots of fitted values appear at each time-point in transparent blue (GEE) and red (LMM), with 95\% confidence intervals (CIs) marked by purple (GEE) and orange (LMM) rectangles. Greater variation and longer, uneven CIs in the imbalanced case highlight the need to assess data balance before applying these methods.
	}\label{fig:balimbal}
\end{figure}

Using synthetic data, we examine sampling balance effects on LMM accuracy by comparing LMM (\texttt{lme4}) and GEE (\texttt{geepack}) in \texttt{R} (\autoref{fig:balimbal}). In imbalanced datasets, linear mean estimation curves at endpoints and fitted value boxplots differ from the balanced case. The imbalance also widens confidence intervals (CIs). In the balanced case (\autoref{fig:balimbal}a), GEE minimizes variance, making boxplots and CIs nearly invisible, whereas precision deteriorates in the imbalanced scenario.

\subsection{Abundances and non-Gaussian relative abundances}\label{GLMMSection}
Gaussian assumption often fails for abundance count data, necessitating a generalization of LMMs. The generalized LMM (GLMM) \citep{Breslow1993} links the expected response to fixed and random effects via a nonlinear function 
\begin{eqnarray}\label{GLMM}
	E[\mathbf{y}_i|\mathbf{b}_i]=g^{-1}(\mathbf{X}_i\boldsymbol{\beta}+\mathbf{Z}_i\mathbf{b}_i),
\end{eqnarray}
where $g^{-1}(\cdot)$ is the inverse link function and $\mathbf{y}|\mathbf{b}$ follows an exponential dispersion family distribution. Despite its suitability for count data, GLMMs remain underutilized in LOD analysis, where normal approximations or transformations are often preferred. GLMM estimation is computationally demanding, but several \texttt{R} packages facilitate implementation, including \texttt{glmm} \citep{Knudson2021}, \texttt{glmmML} \citep{BrostromHolmberg2011}, and \texttt{glmmTMB} \citep{Brooksetal2017} for discrete and zero-inflated data. Bayesian and multivariate approaches can be implemented using \texttt{MCMCglmm} \citep{Hadfield2010} and \texttt{brms} \citep{Burkner2017}.

LOD primarily relies on feature abundances, often transformed into relative abundances. However, normal approximations fail for skewed distributions, and Poisson-based models require high intensity. Overdispersion and zero inflation further challenge normality assumptions, necessitating discrete distributions. Standard GLMM tools, such as \texttt{glmm} \citep{Zhangetal2011} package in \texttt{R}, model binomial and Poisson distributions, while \texttt{Maaslin2} \citep{Mallicketal2021} incorporates compound Poisson and zero-inflated negative binomial models.

A suitable GLMM for longitudinal gut microbiome data is the zero-inflated beta regression model \citep{ChenLi2016}. In a study on pediatric IBD, longitudinal samples were collected from 47 children receiving anti-TNF therapy and 12 undergoing exclusive enteral nutrition (EEN) at baseline, one, four, and eight weeks. Relative abundances at the genus level were analyzed to identify treatment-associated bacterial changes over time. Let $y_{ig}(t_{ij})$ denote the relative abundance of an omics feature $g$ for subject $i$ at time $t_{ij}$ where it is zero with probability $p_{ig}(t_{ij})$ and distributed as $Beta(\mu_{ig}(t_{ij})\phi_g,(1-\mu_{ig}(t_{ij}))\phi_g)$ with probability $p_{ig}(t_{ij})$

The regression models are:
\begin{eqnarray}
	\mathrm{logit}(p_{ig}(t_{ij})) &=& \alpha_0+\mathbf{X}_{ij}\boldsymbol{\alpha}+\varepsilon_{ij}, \label{ChenLi1}\\
	\mathrm{logit}(\mu_{ig}(t_{ij})) &=& \beta_0+\mathbf{Z}_{ij}\boldsymbol{\beta}+\epsilon_{ij}, \label{ChenLi2}
\end{eqnarray}
where $\varepsilon$ and $\epsilon$ are white noise, and $\mathbf{X}$ and $\mathbf{Z}$ denote fixed and random effects. Parameter estimation is performed via maximum likelihood.

\subsection{High-dimensionality of omics data}\label{LMMHigh-dimensionality}
A key challenge in applying LMM to omics data is the high dimensionality due to numerous correlated features. Adaptations, such as the CBCV-CorrConf algorithm \citep{McKennanNicolae2022} (available in \texttt{R} package \href{https://github.com/chrismckennan/CorrConf}{\texttt{CorrConf}}), address this by detecting significant random effects and estimating their influence on correlation structures. This method, validated on sex-specific DNA methylation in a 15 twin-pair study. \citep{Martinoetal2013}, identified 38\% of known sex-associated methylation sites among 330,168 examined, aligning with findings from the longitudinal birth cohort CHAMACOS study \citep{Yousefietal2015}.

Dimension reduction techniques, including tensor factorization \citep{Moretal2022}, have been used in longitudinal proteomics. Alternatively, fitting separate LMMs for each feature, followed by FDR correction, ensures nominal significance, albeit at the cost of testing power. Regularization methods, such as \texttt{glmmLasso} \citep{GrollandTutz2014} and \texttt{glmmixedlasso} \citep{Schelldorferetal2014}, aid in variable selection by minimizing a penalized log-quasi likelihood function. These approaches have been applied to gene expression data to track scarring progression over 11 time-points \citep{Derricketal2020}. Faster alternatives include \texttt{MXM} \citep{Tsagrisetal2018} in R and \texttt{PenalizedGLMM.jl} \citep{StPierreetal2023} in Julia, which accommodate between-individual correlations and binary traits.

An extension of LMM involves evaluating a set of multiple LMMs with predefined variables using multi-model inference and a genetic algorithm \citep{VanderBorghtetal2014}. In a study on RAL resistance due to mutations in the HIV integrase region, 153 subjects with an average of three repeated measurements and 991 clonal genotype-phenotype pairs were analyzed. The study compared OLS and LMM-based variable selection using a genetic algorithm. For discrete omics data, penalized GLMMs extend to ordinal response variables, employing forward stagewise algorithms \citep{HouandArcher2015} to identify predictive gene sets in a longitudinal microarray dataset of inflammation and injury response studies. 

Longitudinal studies often face challenges due to non-random sampling, particularly when costly biomarkers are involved. For example, a biased sampling design \citep{Schildcroutetal2012} was used in a study where daily urine samples determined the optimal day for serum sampling, which coincided with a surge in luteinizing hormone (LH). The study assessed the effect of fiber intake on LH surge probability, introducing bias via an auxiliary variable, modeled as follows
\begin{flalign*}
	\notag&P(\mathrm{LH}_{ij}>20|\mathbf{X}_{ij}, \text{having serum sample for subject }i \text{ at time } j)\\
	\notag&=\mathrm{logit}^{-1}\Big(\log\Big[\\
	&\frac{P(\text{having serum sample for subject }i \text{ at time } j|\mathbf{X}_{ij},\mathrm{LH}_{ij}>20)}{P(\text{having serum sample for subject }i \text{ at time } j|\mathbf{X}_{ij},\mathrm{LH}_{ij}\leq20)}\Big]&\\
	&+\mathbf{X}_{ij}\boldsymbol{\beta}\Big).
\end{flalign*}
A random-effects version of this model for analyzing longitudinal binary data with biased sampling is discussed \citep{Schildcroutetal2012}. The observation-dependent samplings are less expensive, and the probability of inclusion in the sample comes from auxiliary variables. A sequential offset logistic regression model the response variable onto the auxiliary variables and then auxiliary variables onto metadata. This can adjust the imbalance induced by the rarity of a phenotype, e.g., $P(\mathrm{LH}_{ij}>20)\approx0$ or 1, marginally.

\subsection{Differential expression analysis}\label{Differential expression analysis}
Differential gene expression analysis (DEA) identifies genes with varying read counts across conditions. In longitudinal data, the challenge lies in modeling DE dynamics while accounting for repeated measurements. Standard GLMMs must be adapted to capture dependence structures over time.

Some DEA models incorporate this dependence by linking variance to the mean function via a dispersion parameter, particularly in the exponential dispersion family. Although DEA typically applies to gene read counts, it extends to relative abundances with appropriate distribution choices (see Section \ref{DEAAppend}). Most longitudinal studies test fixed effects, often assuming balanced samples and ignoring subject effects, which can inflate true positive rates. However, this issue is minimized in independently generated time-course data.

For example, microbial culture studies treat colony age as a fixed factor, modeling omics feature dynamics accordingly. Wu and Wu \citep{WuWu2013} proposed a nonparametric approach estimating $y_{igc}(t_{ij}) = f_{gc}(t_{ij}) + \varepsilon_{ij}$, testing \begin{eqnarray*}
	\left\{
	\begin{array}{ll}
		H_{g0}:f_{gc}(t)=f_g(t)  & \text{for all }t \text{ and } c,\\
		H_{g1}:f_{gc}(t)\neq f_{gc'}(t)  & \text{for some }c,c',
	\end{array}\right.
\end{eqnarray*} 
for subject $i$, feature $g$, and condition $c$. Rejecting $H_{g0}$ suggests biomarkers, with post hoc tests identifying conditions of variation while controlling FDR.

For balanced time-course data, regular DEA tools treat time-points as conditions. For instance, \texttt{DESeq2} \citep{Love2014} identified gene clusters across post-hatching weeks in killifish \citep{Sacramentoetal2020}. Addressing inter- and intra-subject correlation, Aryee et al. \citep{Aryeeetal2009} modeled log-transformed expression values via multivariate normality, computing posterior probabilities of DE, implemented in \texttt{BETR}. 

While \texttt{limma} \citep{Ritchieetal2015} applies empirical Bayes, factorial designs, or spline regression, it assumes independent observations, making it suitable for time-course but not longitudinal data. For the latter, \texttt{gamm} in \texttt{mgcv} handles generalized additive mixed models \citep{LinZhang1999}, as used in a cohort study profiling transcriptomics, proteomics, metabolomics, cytokines, and growth factors \citep{Sailanietal2020}, revealing 1,133 seasonally varying features linked to cardiovascular health.

A Bayesian DEA model for time-course data \citep{Caoetal2020} assumes negative binomial-distributed omics features:
\begin{eqnarray}\label{NBlinkDEA}
	E[y_{igc}(t)|\mu_{igc}(t)] &=& S_{gc}\exp\Big\{ \theta_{g0}\mathds{1}_{\{c=1\}}\\
	&+& \mathbf{B}^T(t)\boldsymbol{\theta}_{g1}\mathds{1}_{\{c=1\}}+ \mathbf{B}^T(t)\boldsymbol{\eta}_g)\Big\}\nonumber.
\end{eqnarray}
Here, $\mathds{1}$ is an indicator function and $\mathbf{B}^T(t)$ denotes basis functions. The Bayesian method estimates parameters and tests $H_0^g:(\theta_{g0},\boldsymbol{\theta}_{g1}) \neq \mathbf{0}$, implemented in \texttt{MAPTest} \citep{Caoetal2020}. Rejection distinguishes condition $(\theta_{g0}=0)$ and time $(\boldsymbol{\theta}_{g1}=\mathbf{0})$ effects, along with an alternative AR-based test for the time effect \citep{Ohetal2013}.

The \texttt{TiSA} pipeline \citep{Lefoletal2023} integrates \texttt{DESeq2}/\texttt{limma} for DEA in time-course transcriptomics data, clustering, and pathway enrichment via \texttt{goprofiler2}. DEA results appear in heatmaps using recursive thresholding (PART) \citep{Nilsenetal2013}. While PART is not specific to longitudinal data, \texttt{TiSA} adapts it by aggregating time-points. Although some longitudinal omics studies do not explicitly employ Bayesian methods, they utilize normalized counts from \texttt{DESeq2} \citep{Love2014}, which fundamentally relies on Bayesian GLM estimation under a negative binomial distribution. This Bayesian-inspired approach has been applied in single-cell transcriptomics \citep{Tsangetal2017} and previously discussed transcriptomic datasets \citep{Sacramentoetal2020}. A list of tools useful to handle the LOD analyses are provided in Section \ref{AppendBayesian}.

Longitudinal metabolomics DEA \citep{Stanberryetal2013} emphasizes pathways, computing log-relative expression means across time-points. The presented algorithm \citep{Haynesetal2013} identifies high-scoring subpathways, assessing their temporal influence via weighted mean expressions.

\section{Modeling sample paths}\label{Nonparametric approaches}		
\subsection{Functional Data Analysis}
Longitudinal trajectories can be modeled as functions of time or continuous covariates, assuming $y_{ij}\sim N(f(t_{ij}),\sigma^2)$. A common approach is cubic spline estimation with smoothness constraints \citep{Metwallyetal2022}, facilitating visualization and hypothesis testing for group differences (e.g., case vs. control). This requires sufficient repeated measurements per subject. The \href{https://bioconductor.org/packages/3.16/bioc/html/OmicsLonDA.html}{\texttt{OmicsLonDA}} package provides an accessible implementation for LOD analysis.

A stochastic mixed model incorporates dependence structures and metadata effects:
\begin{eqnarray*}
	y_{ij}=X_i\boldsymbol{\beta}+f(t_{ij})+b_i+U_i(t_{ij})+\varepsilon_{ij},
\end{eqnarray*}
where $f$ is a smooth function of time, $b_i\sim N(0,\phi)$ captures subject-specific intercepts, and $U_i(t)$ is a Gaussian process with correlation function $\eta_{\rho}(t,s)\equiv \mathrm{Corr}(U_i(t),U_i(s))$ for the correlation parameter $\rho$. This model has been used to study progesterone levels across menstrual cycles while adjusting for BMI and age \citep{Zhangetal2000}. A simplified functional regression model removes random effects \citep{Meyeretal2007}, applied in the Study of Women’s Health Across the Nation (SWAN) dataset to analyze menstrual cycle hormone levels with B-splines, revealing inflection points at the luteal transition \citep{RamsaySilverman2005}.

Another FDA approach models longitudinal phenotypes, $y$, against genetic markers \citep{MarchettiBowick2016}:
\begin{eqnarray*}
	y_{ij} = f_0(t_{ij})+\sum_{g=1}^G\sum_{\gamma=0}^2f_g^{\gamma}(t_{ij})X_g^{\gamma}+ \varepsilon_{ij},
\end{eqnarray*}
where $f_0$ is the intercept function, $f_g^{\gamma}$ represents genotype-specific genetic effects, and $G$ denotes the number of genes. This high-dimensional (only one of the terms$f_g^0X_g^0+f_g^1X_g^1+f_g^2X_g^2$ is nonzero per observation) genome-wide association studies (GWAS) model is implemented in \texttt{Time-Varying Group SpAM} code in \texttt{Matlab} \citep{MarchettiBowick2016} and has been applied to GWAS asthma studies.

Interdomain Gaussian Process (GP) modeling treats time-dependent responses as a GP realized at sparse points, inferring posterior distributions across domains. Formally, for subject $i$:
\begin{eqnarray*}
	\mathbf{y}_i|X_i,\mathbf{t}_i\overset{ind}{\sim}N(f(X_i),\sigma^2),\quad f\sim\mathcal{G}(0,k(\cdot,\cdot)),
\end{eqnarray*}
where $k$ is a kernel function capturing intra-subject dependence, estimated using predefined structures. \texttt{LonGP} \citep{Chengetal2019} extends this by decomposing $f$ into additive components with different kernels, supporting various covariate types. The provided \texttt{Matlab}-based code supports squared exponential and periodic kernels for continuous covariates, as well as constant, binary, and categorical kernels for factors. For model selection, \href{https://github.com/omicsEye/waveome}{\texttt{waveome}} \citep{Rossetal2022} employs BIC with a more flexible model that includes both summation and multiplication of kernel functions, while \texttt{LonGP} uses leave-one-out and stratified cross-validation with Bayesian bootstrap.

\texttt{LonGP} analyzed 758 stool samples from 222 children (birth to age three) located in three countries to assess socioeconomic effects on the gut microbiome \citep{Chengetal2019}, identifying reduced microbial enrichment in Russian children \citep{Vatanen2016}. Applied to plasma proteomics from T1D and control children, it identified 38 disease-associated proteins, including 18 missed by LMM \citep{Liu2018}. While \texttt{LonGP} assumes normal and Poisson models, \texttt{GPcounts} extends it to negative binomial and zero-inflated models \citep{BinTayyashetal2021}.

\subsection{Low/high replication}
Longitudinal studies leverage high replication and regularity to mitigate sparsity issues \citep{Vaas2012}. Phenotype microarray data, used to characterize microbial growth under specific conditions, offers balanced measurements but poses computational challenges for LMMs due to high-dimensional covariance structures. Curve fitting methods, such as cubic splines via \texttt{smooth.spline()} in \texttt{R}, provide a computationally efficient alternative. A study on 2 microbial species (\textit{E. coli} and \textit{P. aeruginosa}) $\times$ 4 strains $\times$ 2 biological replicates $\times$ ten technical replicates $\times$ 96 substrates provided 7,680 time-dependent measurements and instead of comparing model parameters, the estimated area under the curves (AUC) were compared as a growth summary \citep{Vaas2012}. However, \texttt{OmniLog}\textsuperscript{\tiny\textregistered} software introduces estimation biases, neglecting part of the stored information.

For LOD with small sample sizes, frequentist methods are not recommended. However, Bayesian GLMMs \citep{Fongetal2009}, Kenward-Roger approximation (\href{https://CRAN.R-project.org/package=pbkrtest}{\texttt{pbkrtest}} \citep{HalekohHojsgaard2014}), and multilevel modeling \citep{McNeish2017} remain viable approaches.

\section{Longitudinal omics: few subjects, many features}\label{Single-cell}
Longitudinal omics, such as single-cell (sc) analysis, lags behind other omics due to limited samples and high-dimensional measurements. scRNA samples, typically from the same tissue, track treatment response or cancer evolution \citep{Sharmaetal2018}. The fishplot method \citep{Milleretal2016} visualizes cell population proportions over time but loses symmetry when plotting clonal frequencies. Some scDNA approaches adapt bulk sequencing methods. A study of 47 metastatic patients modeled detectable and undetectable mutations at each time point \citep{Khanetal2018}, using a two-stage mixed model for changes from baseline and a Cox model for survival analysis.

Another approach characterizes disease progression via candidate driver mutations \citep{Ramazzottietal2022}. A time-dependent random matrix represents somatic mutations, accounting for false positives, false negatives, and missing data. Boolean matrix factorization estimates cell attachment and phylogenetic matrices, maximizing a likelihood function via MCMC. The longitudinal analysis of cancer evolution (LACE) algorithm, available in \texttt{R} package \href{https://github.com/BIMIB-DISCo/LACE}{\texttt{LACE}}, tracks genotype prevalence and reconstructs clonal evolution. Unlike \texttt{CALDER} \citep{Myersetal2019}, which clusters mutations based on allele frequencies, \texttt{LACE} controls error rates via weighted likelihood maximization, incorporating uncertainty through sum-condition-based factorization \citep{ElKebiretal2015}.

The \texttt{PALMO} pipeline enables longitudinal multi-omics analysis, spanning single-cell and bulk data \citep{Vasaikaretal2023}. It includes: (1) Variance decomposition analysis (VDA), using LMM to partition variation; (2) Coefficient of variation profiling (CVP), distinguishing stable vs. oscillating bulk features; (3) Stability pattern evaluation across cell types (SPECT), the scRNA counterpart of CVP; (4) Outlier detection analysis (ODA), identifying weak intra/inter-donor correlations; and (5) Time course analysis (TCA), using \texttt{MAST} for differential expression analysis. \texttt{PALMO} also provides visualizations, including circos plots, CV heatmaps, and UMAP projections. The study analyzed 60 plasma and PBMC samples from six donors over 10 weeks, integrating plasma protein abundances, flow cytometry, scRNA-seq (24 samples from four donors), and scATAC-seq (18 of these 24). \texttt{Seurat} identified 31 cell types from 472,464 cells, with 11,191 highly expressed genes. VDA, treating donors and time points as random effects, revealed strong inter-donor variation in CBC, PBMC, and plasma proteins, identifying 75 proteins with greater intra- than inter-donor variability. Including cell type as a random effect in single-cell data showed stronger inter-cell type variation. CVP identified 413 longitudinal and 629 stable proteins, the latter as potential biomarkers. ODA flagged weak intra-donor correlations at week 6 in one subject. SPECT counted gene CV exceedances across 76 combinations of 4 donors and 19 cell types, detecting 700 super-variable, 2,129 super-stable, 5,750 variable, and 4,004 stable (STATIC) genes, with STATIC genes as biomarker candidates.

\section{Classification based on longitudinal omics data}\label{Classification}
Logistic regressions using GLMM are commonly employed for classifying longitudinal omics data. For instance, a GLMM for ordinal data classified 657 buffy coat samples from burn injury patients hybridized to Affymetrix Human Genome U133 Plus 2.0 Arrays, identifying key genes associated with organ failure  \citep{HouandArcher2015}.

Linear Discriminant Analysis (LDA) is widely used in omics classification but struggles with time effects and imbalanced case-control ratios in rare disease studies. To address these, Longitudinal Discriminant Analysis \citep{Marshall2000} (LonDA) estimates group-specific parameter distributions over time, defining decision boundaries as:
\begin{eqnarray}\label{ESL:LDA} 
	&&\log\pi_k +\log f_k(\mathbf{y}_i;\phi_k(\mathbf{t})) \\
	&&~~~~=\max_j \{\log\pi_j +\log f_j(\mathbf{y}_i;\phi_j(\mathbf{t}))\}, ~j=1,\ldots,g. \nonumber
\end{eqnarray}
With Gaussian group densities and equal covariances, the boundaries remain linear. LonDA accounts for mean and variance heterogeneity across time, leading to four estimation scenarios \citep{Marshall2000} (see Section \ref{AppenLonDA}): homoscedastic, mean-heteroscedastic, variance-heteroscedastic, and fully heteroscedastic models. The model has been applied to classify pregnancy outcomes based on beta-subunit measurements  \citep{Marshall2000} and later analyzed via Bayesian nonlinear classification  \citep{delaCruzMesia2007}, multivariate clustering  \citep{Villarroeletal2009}, and bivariate modeling under missing data assumptions  \citep{Marshall2009}, emphasizing the need for time-aware classification.

The Quadratic Inference Function Classifier \citep{WangQu2014} (QIFC) efficiently handles small sample sizes and, unlike LonDA, omits full covariance estimation by modeling time dependencies via a QIF-based distance. After fitting a semiparametric model, it computes the distance of a data-point from each class and then assigns it to the closest. Applied to omics datasets, QIFC outperformed logistic regression, SVMs, and functional data classifiers, achieving a $5.3\%$ misclassification rate on yeast (\textit{Saccharomyces cerevisiae}) cell cycle data \citep{Spellmanetal1998} (2467 genes, 79 time-points) and $14\%$ on \textit{Drosophila melanogaster} data \citep{Arbeitmanetal2002} (4028 genes, 70 time-points).

\begin{figure*}[!ht] 
	\centering 
	\includegraphics[width=.9\textwidth]{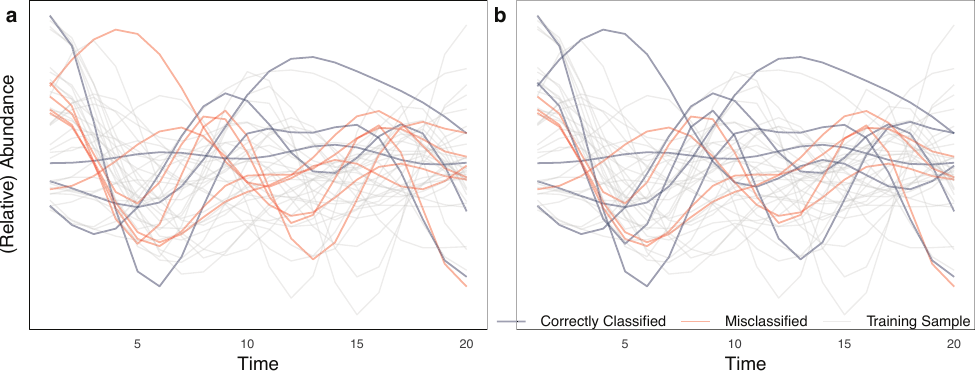} 
	\caption{Comparison of \textbf{a} logistic regression and \textbf{b} Bayesian classifier \citep{Daietal2017} using $m=30$ subjects at $n_i=20$ time-points, where 20 subjects were used for training and 10 for testing. Red lines indicate misclassified, and blue lines correctly classified samples.}\label{fig:Classification}
\end{figure*}

FDA optimal Bayes classifier \citep{Daietal2017} offers a nonparametric classification approach by projecting time-dependent data onto specific directions (e.g., principal components). Applied to yeast data, this method achieved a $12.5\%$ misclassification rate, outperforming logistic regression and centroid-based classifiers \citep{DelaigleHall2012}. Figure \ref{fig:Classification} compares logistic regression and the Bayesian functional classifier on sample data.

Regularization methods address limitations of LDA in high-dimensional cases by estimating class-dedicated functions of time:
\begin{eqnarray*} 
	\mathbf{y}_i = \mathbf{X}_i \boldsymbol{\beta} + \mathbf{f}_k(\mathbf{t}_i) + \boldsymbol{\varepsilon}_i. 
\end{eqnarray*} 
Here, the nonparametric lasso estimate of $\mathbf{f}_k$ is provided by a penalized EM algorithm \citep{Arribas-Giletal2015}.

Functional principal component (FPC) analysis has been used to classify Alzheimer’s patients based on 319 longitudinal biomarkers \citep{LiandHsu2022}. The FPC scores of subject $i$, $\mathbf{x}i$ is used to minimize the loss function
\begin{eqnarray*} 
	l(\boldsymbol{\beta}) = \frac{1}{N} \sum_{i=1}^m e^{-c_i \mathbf{x}i^T \boldsymbol{\beta}} + \lambda \sum_{k=1}^K \sqrt{p_k} \|\boldsymbol{\beta}_k\|_2, 
\end{eqnarray*} 
where $c_i$ is a binary disease class indicator, $\boldsymbol{\beta}_k$ is the coefficient corresponds to $k^{\mathrm{th}}$ feature group, and $p_k$ is the number of FPC scores.

A distribution-free method \citep{Pfeifferetal2012} integrates longitudinal biomarkers via a composite marker score, ensuring no information loss in regression through sufficient dimension reduction. The method assumes decomposability of marker vectors into marker and time effects and models the response variable using a sufficient linear combination of markers, accounting for right censoring. Applied to a cancer cohort of 100 cases and 100 controls observed at 4 time-points, logistic-type regression classified cases based on uric acid, blood glucose, and serum cholesterol measured at four time-points.

Random forest (RF)-based classifiers are also widely used. RF++ \citep{Karpievitchetal2009} library in \texttt{C++} employs subject-level averaging and bootstrapping, successfully classifying 38 subjects (30 cancer cases and 8 controls), with substantial replication per subject. Despite subject-level averaging, it does not capture time effects  \citep{Hajjemetal2014}. Repeated measures RF has been applied to classify diabetes patients’ observation times into high-risk nocturnal hypoglycemia based on metabolomics data  \citep{Calhounetal2021}.

\section{Multivariate outcomes in clinical longitudinal studies}\label{Multivariate outcomes}
Omic data involves multiple correlated features measured simultaneously. While univariate analyses with multiple testing corrections (e.g., B-H) usually suffice \citep{VerbekeetalReview2014}, they fail to capture inter-feature dependencies, leading to high-dimensional GLMM challenges (see Section \ref{LMMHigh-dimensionality}). Multivariate models offer a more holistic approach, particularly for the analysis of known biomarkers.

Consider $p$ outcomes measured at $n_i$ time points for subject $i=1,\ldots,N$, represented as $Y_i=[\mathbf{y}_{i1}|\dots|\mathbf{y}_{ir}]$, an $n_i\times r$ matrix. The full dataset forms an $N \times r \times n_i$ array, with measurement errors $\mathcal{E}_i=[\boldsymbol{\varepsilon}_{i1}|\dots|\boldsymbol{\varepsilon}_{ir}]$.

The multivariate $t$ linear mixed model (MtLMM) models the vectorized response $\mathbf{y}_i = \text{vec}(Y_i)$ as:
\begin{equation*}
	\mathbf{y}_i=X_i\boldsymbol{\beta}+ Z_i\mathbf{b}_i+\boldsymbol{\mathcal{E}}_i,
\end{equation*}
where block-diagonal matrices $X_i$ and $Z_i$ allow different covariate sets per outcome. Measurement errors follow a $q_.+n_i$-variate central $t$ distribution with scale matrix $\text{diag}\{D,R_i\}$ and $\nu$ degrees of freedom. The covariance structure $R_i=\Sigma_{\text{Err}} \otimes C_i$ accommodates inter- and intra-subject correlations via a damped exponential correlation (DEC) function. Parameter estimation under missing at and not at random employs the alternating expectation-conditional maximization algorithm \citep{WangMtLMM2013} with Fisher scoring, and empirical Bayes shrinkage \citep{VerbekeMolenberghs2009} estimates random effects.

Data from 161 women with 124 normal and 37 abnormal deliveries tracked estradiol and $\beta$-HCG levels during early pregnancy. Log-transformed responses were modeled with covariates (rescaled time, squared time, pregnancy group). AIC favored MtLMM with DEC, while BIC preferred AR(1), both dominated the normal LMM. MtLMM variants have also been extended to censored heavy-tailed \citep{Wangetal2018} and missing data scenarios \citep{LinWang2022,Wang2020}.

The MtLMM was also applied to HIV-1 RNA counts in seminal and blood plasma from 149 subjects (106 on therapy, 43 not) with DEC, outperforming normal LMM. Penalized likelihood extensions addressed high-dimensionality \citep{TaavoniArashi2022}. 

A constrained multivariate LMM analyzed BioCycle data \citep{Howardsetal2009} (259 women, ages 18–44), incorporating ground truth constraints (e.g., EC peaks before LH surge, followed by PG rise). The model for hormone $h$ is
\begin{equation*}
	y_{ijh}= \mu_{jh}+\mathbf{X}_{ij}^T\boldsymbol{\beta}_h+\xi_{ih}+\zeta_{ij}+\varepsilon_{ijh},
\end{equation*}
enforces $\mathbf{A}\boldsymbol{\mu} \leq \mathbf{b}$ for structured hormone progression where $\mu_{jh}$ is the marginal mean, $\xi_{ih}$ is the subject-specific random effect, and $\zeta_{ij}$ is the time effect. The constraint decreases bias by limiting the parameter space. An adaptive MCMC estimates posterior summaries \citep{Royetal2012}.

\section{Other Scenarios}\label{Other scenarios}
Some approaches do not fit neatly into one category but remain relevant to previous sections. We discuss them here.

\paragraph{Integration}
High-dimensionality is the key challenge in the integration of inferences based on multiple omics features from the same subject. For instance, in a study classifying subjects into type 2 diabetes and prediabetes using longitudinal transcriptomes, metabolomes, cytokines, and proteomes, intra-correlations were ignored by subtracting the mean of healthy time points from baseline measurements \citep{Zhouetal2019}. While this avoids estimating correlation effects, it sacrifices data granularity by summarizing with AUC values.

Longitudinal multi-omics data, particularly with small sample sizes are often analyzed feature-by-feature  \citep{Bernardesetal2020}. However, when data tables can be merged, high-dimensional methods are used, such as in studies of gut microbiome, host transcriptome, and methylome in IBS patients  \citep{Marsetal2020}. 

The Bayesian Additive Regression Trees (BART) method integrates multi-omics data in longitudinal studies via a two-stage modeling approach, as applied to the Integrative Human Microbiome Project  \citep{Mallicketal2024}. To integrate longitudinal DNA methylation, mRNA, miRNA, and proteomics from developing murine alveoli,  \citep{Dingetal2019} introduced iDREM, which identifies transcription factors and miRNAs influencing gene expression via an input-output hidden Markov model combined with logistic regression  \citep{Schulzetal2013}.

\paragraph{Clustering and abundance change network modeling}

Clustering longitudinal omics data remains an emerging field, with existing methods yet to be fully adapted  \citep{DenTeulingetal2023,Zhouetal2023}. However,  \citep{Larsenetal2022} employed the clustering approach of  \citep{Genolinietal2016} to classify patients based on breast cancer LOD.

\paragraph{ImpulseDEA}

Longitudinal RNA and DNA methylome data from IBD patients were analyzed across two cohorts receiving anti-TNF therapy (infliximab or adalimumab) and a control cohort on vedolizumab  \citep{Mishraetal2022}. Subjects were fully observed over time, with both pairwise time-point comparisons and longitudinal approaches used to identify gene expression differences. The analysis employed the ImpulseDE2 model  \citep{Fischer2018}.

\paragraph{Survival and dropuot}
In omics studies, particularly those involving severe diseases, subjects may leave before the study ends due to treatment changes, withdrawal, recovery, or death. When the primary focus is the time of exit (e.g., recovery/mortality), survival models are employed. While survival data is often sparse, some studies integrate longitudinal measurements to predict outcomes, such as CD4 counts in AIDS cohorts \citep{Tsiatis1995,WangTaylor2001} and PSA levels in prostate cancer \citep{Yuetal2004}. Cox regression remains the dominant approach, either independently or within joint models that assess the impact of longitudinal omics and baseline metadata on hazard rates \citep{Sacramentoetal2020}.

A key advancement in joint modeling combines LMM with Cox regression to capture disease progression and dropout mechanisms. For instance, in a Zidovudine (ZDV) trial for HIV, CD4 counts served as surrogate markers for survival, with empirical Bayes estimates adjusting for missing data \citep{Tsiatis1995}. Extensions incorporated Bayesian inference and Ornstein-Uhlenbeck processes for time-varying coefficients, refining survival estimates \citep{WangTaylor2001}. Change-point models have also been applied to plasma HIV RNA data from AIDS trials, capturing virologic failure dynamics and dropout effects using Bayesian MCMC methods  \citep{Ghosh2011}. Given the predominant use of survival models in time-to-event analyses rather than longitudinal omics, a detailed discussion of these methods is provided in the supplementary material. More discussion in this topic is available in Section \ref{AppenSurvival}.

Finally, the topic of causality is of general interest in omics studies, where the goal is to transition from statements of association to causal effects. One line of research that attempts to address this is Mendelian Randomization (MR) \citep{Sandersonetal2022}. A quasi-experimental study \citep{RichmondSmith2022} was designed using instrumental variables in the form of single-nucleotide polymorphisms as a way to estimate the causal effect of a biomarker of interest, usually through variations of two-stage linear models, potentially with random effects. One may use both MR and LMMs to estimate the causal effect of metabolite biomarkers in a longitudinal cohort \citep{Porcuetaletal2021}. While another used functional data analysis methods To account for time-varying exposures when estimating the effect of biomarkers on outcomes of interest, functional data analysis can be used \citep{Caoetal2016}. Since \href{https://cran.r-project.org/web/packages/MendelianRandomization/index.html}{\texttt{MendelianRandomization}} and \href{https://cran.r-project.org/web/packages/MendelianRandomization/index.html}{\texttt{TwoSampleMR}} packages to implement MR currently only focus on cross-sectional data, extending these methods to longitudinal contexts is an active area of research.

\section{Discussion}
We highlighted challenges in LOD analysis, where classic methods like LMM and GLMM struggle with intra-subject correlation, high-dimensionality, irregular sampling, and normality assumption. Nonparametric solutions for correlation structures remain underdeveloped, and imputation techniques require improvements to better preserve biological variability.

Multivariate methods are particularly limited when handling high-dimensional omics data, as most approaches fail beyond 10–20 response variables \citep{WangLou2017, VerbekeetalReview2014}. The redundancy of certain genes at specific time points further complicates analysis. Nonparametric FDA methods offer potential solutions but remain computationally demanding. Additionally, log-transformation normality assumptions often fail in real datasets, and Bayesian extensions for LMM/GLMM in high-dimensional settings are lacking. Approaches such as joint modeling of survival and longitudinal data improves efficiency, reduces bias, and enable dynamic predictions. While many software tools exist, their scalability is uncertain, as most were developed for specific datasets and require regularization for broader applications. Future research should focus on computationally efficient and adaptable methods for high-dimensional longitudinal omics data.

We summarized key methods for longitudinal omics analysis and noted available software tools, highlighting their strengths and limitations. Scalable and flexible models are still needed to handle high-dimensional, irregular, and sparse data. Combining nonparametric, Bayesian, and deep learning approaches will be key to future progress.

\bibliographystyle{plainnat}
\bibliography{References}

\newpage
\section*{Supplementary materials}
\setcounter{section}{0}

\renewcommand{\thesection}{Supplementary \Alph{section}}

\makeatletter
\renewcommand{\l@section}{\@dottedtocline{1}{1.5em}{2.3em}}
\renewcommand{\numberline}[1]{#1\quad}
\renewcommand{\@seccntformat}[1]{\csname the#1\endcsname.\quad}
\makeatother

\renewcommand{\thefigure}{S\arabic{figure}}
\renewcommand{\theequation}{S.\arabic{equation}}
\setcounter{figure}{0}
\setcounter{equation}{0}

The \texttt{Categorized References.csv} file summarizes the entry documents. The 0/1 values indicate whether an article addresses—or whether the mentioned methodology can address—the subject listed in each column title. The last three columns specify whether the proposed methodology in the entry document \textit{can} potentially be used for genomic data, differential expression analysis (DEA), or time-course data. Algorithms \ref{alg;overall} and \ref{alg;FDR} help break down the study's scientific question into general components and point to the relevant sections of the manuscript that address each one.

	\begin{algorithm}
		\caption{Overall steps in the analysis of LOD}\label{alg;overall}
		\begin{algorithmic}[1]
			\Require Omics and metadata tables
			\If{number of omics features of interest $< 10$ \citep{VerbekeetalReview2014}}
			\If{Multivariate normal or $t$ distribution assumption is fulfilled}
			\State Choose one of the methods discussed in Section~\ref{Multivariate outcomes}, considering the balance of the sampling design.
			\Else
			\State Implement Algorithm~\ref{alg;FDR}
			\EndIf
			\Else
			\If{The question is ``Are the omics features affected by the metadata?"}
			\State Implement Algorithm~\ref{alg;FDR}
			\Else
			\If{The scale of the metadata of interest is count or continuous}
			\If{Normality assumption is fulfilled for the biomarker measurements}
			\State Use high-dimensional LMM with or without survival analysis (depending on the study design), considering the balance of the sampling design.
			\Else
			\State Use the corresponding high-dimensional GLMM depending on the distribution of biomarker measurements, with or without survival analysis (depending on the study design), considering the balance of the sampling design.
			\EndIf
			\Else
			\State Depending on the sampling design balance, use one of the methods reviewed in Section~\ref{Classification}, or apply joint modeling of GLMM across survival as discussed in Section~\ref{AppenSurvival}.
			\EndIf
			\EndIf
			\EndIf
		\end{algorithmic}
	\end{algorithm}
	
	\begin{algorithm}
		\caption{Univariate FDR Procedure}
		\label{alg;FDR}
		\begin{algorithmic}[1]
			\Require Omics and metadata tables
			
			\For{each omics feature}
			\If{Normality assumption is fulfilled}
			\State Consider the sparsity of the data and the balance of the sampling design
			\If{Data is not survival}
			\State Fit one of the LMM-based methods discussed in Section \ref{continuous}
			\Else
			\State Fit one of the joint models in Section~\ref{AppenSurvival}
			\EndIf
			\Else
			\If{A non-Gaussian underlying distribution exists with a suitable link function that models the relationship between the average value of the omics feature and a linear function of metadata}
			\State Fit one of the GLMM-based methods discussed in Section~\ref{GLMMSection}
			\Else
			\State Estimate the relationship between the average value of the omics feature and metadata nonparametrically
			\State Compute the significance of each covariate using the method in \cite{Lavergneetal2015}
			\EndIf
			\EndIf
			\State Save the corresponding $p$-values for the associated coefficients
			\EndFor
			
			\State Compute adjusted $p$-values using procedures such as Benjamini-Hochberg
			\State Apply a threshold to the adjusted $p$-values
			\State Determine significantly associated features and covariates
			
		\end{algorithmic}
	\end{algorithm}

\section{Linear and generalized linear mixed effects models}\label{AppenLMMGLMM}
Let $\mathbf{y}_i$ be the $n_i \times 1$ vector of all measured values (such as count, abundance, relative abundance, intensity, etc.) from the $i^\text{th}$ subject (patient) for $i = 1, \ldots, m$. The metadata provides the $n_i\times p$ design matrix of the fixed effect of $i^\text{th}$ subject denoted by $\mathbf{X}_i$. Gene expressions and microbiome information of each subject, like a fingerprint, are uniquely dedicated to that subject, and thus, multiple measurements from the same subject produce intra-subject correlation, which must be included in the models. The dependence structure of the multiple measurements out of an individual is explained by unknown individual effect, $\mathbf{b}_i$ with known $n_i\times k$ design matrix $\mathbf{Z}_i$ which links the unknown \textit{random effect} $\mathbf{b}_i$ to the observed measurements $\mathbf{y}_i$ by 
\begin{eqnarray}\label{AppenLMM}
	\mathbf{y}_i=\mathbf{X}_i\boldsymbol{\beta}
	+\mathbf{Z}_i\mathbf{b}_i+ \boldsymbol{\varepsilon}_i,
\end{eqnarray}
for $i=1,\ldots,m$ where $\boldsymbol{\varepsilon}_i\overset{iid}{\sim} N(\mathbf{0},\boldsymbol{\Sigma}_{\mathrm{Err}})$ is Gaussian white noise independent of the Gaussian random effects $\mathbf{b}_i \overset{iid}{\sim}N(\mathbf{0},\boldsymbol{\Sigma}_{\mathrm{rndEff}})$ and in this way
\begin{eqnarray}\label{AppenLMMVar}
	\mathrm{Var}(\mathbf{y}_i)=\boldsymbol{\Sigma}_{\mathrm{Err}}+\mathbf{Z}_i\boldsymbol{\Sigma}_{\mathrm{rndEff}}\mathbf{Z}_i^T,
\end{eqnarray}
which reflects the intra-subject dependence in the second term. Eliminating this correlation from the analysis is equivalent to treating the random effect as a constant effect ($\Sigma_{\mathrm{rndEff}} = \mathbf{0}$), which leads to incorrect inference about $\boldsymbol{\beta}$ and underestimates the variance of the estimates. 
In the imbalanced case, the time effect is often incorporated into $\mathbf{Z}$; that is, its columns represent the multiplication of time or the squared value of time by the fixed effects. This leads to a nonstationary covariance structure for $\mathbf{y}_i$ even within a subject, meaning that the trajectory $y_{i1},\ldots,y_{in_i}$ observed as responses of subject $i$ at times $t_{i1},\ldots,t_{in_i}$, is a realization of a nonstationary process, regardless of the mean function. A better understanding of this analysis can be gained from a two-stage modeling approach: first, fitting a linear model between each subject's measurements and the time-related components (intercept, time, square value of time, etc.), and then regressing the $m$ groups of resulting coefficients onto the fixed effects. The maximum likelihood estimator for the LMM described above has a considerable computational cost. Therefore, an efficient EM algorithm \citep{LairdWare} is usually hired to compute the restricted maximum likelihood estimates.

Gaussianity is a very restrictive assumption for abundance count data, necessitating a generalization of the LMM. The GLMM \citep{Breslow1993} parameterizes the conditional expectation of the response variable as   
\begin{eqnarray}\label{AppenGLMM}
	E[\mathbf{y}_i|\mathbf{b}_i]=g^{-1}(\mathbf{X}_i\boldsymbol{\beta}+\mathbf{Z}_i\mathbf{b}_i),
\end{eqnarray}
where $g^{-1}(\cdot)$ is the inverse link function and $\mathbf{Y}|\mathbf{b}$ belongs to the exponential dispersion family. Despite the flexibility and covering a wide spectrum of distributions, including count and ordinal data, this model has often not been applied to the analysis of LOD. Instead, researchers frequently either ignore the proper likelihood assumptions or use a concave transformation, such as $\log(\cdot)$, to use the normal approximation. Computing the GLMM estimates, whether using frequentist or Bayesian approaches, remains challenging. Researchers interested in using GLMM for their longitudinal studies can employ \texttt{glmm} \citep{Knudson2021}, \texttt{glmmML} (for Poisson and binomial responses) \texttt{R} packages for univariate continuous or discrete response data and \texttt{glmmTMB} \citep{Brooksetal2017} for zero-inflated features. For multivariate analysis, \texttt{MCMCglmm} \citep{Hadfield2010} and \texttt{brms} \citep{Burkner2017} are recommended, with \texttt{brms} specifically considering the survival data. The last two packages are implemented within a Bayesian statistical framework.

Model \eqref{GLMM} is also used to address a different class of models involving time-varying covariates. Here, we incorporate the effect of time as a ratio scale to highlight the dependency of the covariate $\mathbf{X}^T=(\mathbf{X}_{it_{i1}}|\ldots|\mathbf{X}_{it_{in_i}})$ where $\mathbf{X}_{it_{ij}}^T$ is the row vector of covariates for the $i^\text{th}$ subject at time $t_{ij}$. Note that the covariates are not only random but also constitute a stochastic process. This aspect is often overlooked in the analysis of LOD, where metadata are typically treated as fixed effects without accounting for their random structure. While this approach does not introduce additional bias into the ordinary least squares (OLS) estimates of the parameters, it does affect the significance of test results. According to the basic types of time-varying covariates \citep{LaiSmall2007}, two well-known types of modeling are subject-specific and population-averaged models. The subject-specific model is a specific version of the GLMM, considering the subject effect as one of the random effects given by
\begin{eqnarray*}
	y_{it_{ij}}|s_i\sim\mathcal{L}(\mu(\mathbf{X}_{it_{ij}},\mathbf{Z}_i)),~~~\mathrm{where}~~~
	s_i\overset{iid}{\sim}F_s,
\end{eqnarray*}
where $\mathcal{L}$ denotes the distribution of response with mean $\mu(\mathbf{X}_{it_{ij}},\mathbf{Z}_i)$ and $F_s$ is the distribution of the random subject effect $s_i$. The systematic part is formulated using a link function as in \eqref{GLMM}. In population-averaged models, the response variable is linked to the subject's time-varying behavior via the mean function rather than the process $\{\mathbf{x}_{it}\}_t$. Specifically, 
\begin{eqnarray*}
	y_{it_{ij}}|s_i\sim\mathcal{L}(\mu(\mathbf{X}_{it_{ij}}), \phi V(\mu(\mathbf{X}_{it_{ij}}))),
\end{eqnarray*} 
where $V(\cdot)$ is the variance function when the process $\{\mathbf{x}_{it}\}_t$ is nonstationary, and the distribution $\mathcal{L}$ belongs to exponential dispersion family. The conditional mean of the distribution is linked to the covariates and random effects. The generalized estimating equations (GEE) available via \texttt{R} package \texttt{geepack} \citep{Halekohetal2006} or generalized method of moments (GMM) are commonly used for parameter estimation in time-dependent models. 

We bring the use of LMMs or GLMMs with time-varying factors in longitudinal omics studies into attention because these models typically assume that covariates have been fixed (baseline values). When the value of a feature in the metadata depends on its previous value or outcome at an earlier time-point, it becomes an endogenous factor, which can cause bias in estimating metadata effects. This important and widespread issue was discussed by Qi et al. \citep{Qianetal2020} in the literature and further elaborated in a rejoinder version \citep{Qianetal2020Rejoinder}. They proposed a mechanism that, under a conditional independence assumption, allows the use of standard software for fitting LMMs in the presence of endogenous factors. Without this approach, studies would need to develop their methodologies to control for the bias introduced by endogenous factors. 

\section{DEA from hypothesis testing viewpoint}\label{DEAAppend}
DEA in longitudinal studies aims to identify features that change across conditions and/or over time.
Let $\mathbf{y}_{ig}$ be the vector of all read counts for an omics feature (such as a gene) $g$ from subject $i$, and suppose that $\mathbf{y}_{ig}\sim p(\cdot|\boldsymbol{\theta}_g ,\boldsymbol{\eta}_g)$ where $\boldsymbol{\theta}_g$ is the parameter vector that captures changes over time and $\boldsymbol{\eta}_g$ represents nuisance parameters. GLMM can be useful for incorporating the random effects of subjects by replacing $\boldsymbol{\theta}_g$ with $\boldsymbol{\theta}_{ig}$. 
The primary focus in DEA is hypothesis testing on $\boldsymbol{\theta}$ rather than estimation or prediction. In this context, hypothesis testing involves simultaneous testing, where concepts such as the false discovery rate (FDR) or the expected false positive rate become more relevant than the traditional type I error. 
If the hypotheses take the form $H_0^g:\boldsymbol{\theta}_g\in\Theta_0$ versus $H_1^g:\boldsymbol{\theta}_g\in\Theta_1$, a widely used approach is multiple Bayesian testing based on the Bayes factor 
\begin{eqnarray}\label{generalDEtestStatistic}
	\delta(\mathbf{y})= \frac{Pr(\boldsymbol{\theta}_g\in\Theta_0)\times Pr(\mathbf{y}\in\mathcal{R})}{Pr(\boldsymbol{\theta}_g\in\Theta_1)\times Pr(\mathbf{y}\notin\mathcal{R})},
\end{eqnarray}
where $\mathcal{R}=\{\mathbf{y}:(\sum_{g=1}^{G_0}p(\mathbf{y}|\boldsymbol{\theta}_g,\boldsymbol{\eta}_g))(\sum_{g=G_0+1}^{G}p(\mathbf{y}|\boldsymbol{\theta}_g,\boldsymbol{\eta}_g))^{-1}\leq s\}$, is the rejection region, $G_0$ is the number of true nulls and $G$ is the total number of genes. For instance, consider the case of two time-points, $t_1$ and $t_2$ and two conditions $c_1$ and $c_2$. Suppose $\boldsymbol{\theta}_g$ is of the form $(\theta_{g11},\theta_{g12},\theta_{g21},\theta_{g22})$, where $\theta_{gct}$ represents the mean of the distribution of the omics feature under condition $c$ at time-point $t$. Then, $H_0^g:\boldsymbol{\theta}_g\in\Theta_0=\{\boldsymbol{\theta}_g: \theta_{g11}+\theta_{g12}=\theta_{g21}+\theta_{g22}\}$ tests the effect of condition, while $H_0^g:\boldsymbol{\theta}_g\in\Theta_0=\{\boldsymbol{\theta}_g: \theta_{g11}+\theta_{g21}=\theta_{g12}+\theta_{g22}\}$ tests the effect of time. For large enough $G_0$, $G$ and $G-G_0$, the Bayes factor does not depend on $G_0$. Hwang and Liu \citep{HwangLiu2010} demonstrated that the value of 
$s$ can be calculated to keep the FDR below a nominal level, and based on the Neyman-Pearson lemma, it maximizes the average power. 

One straightforward method in DEA for LOD is to fit an LMM or GLMM, and then apply an appropriate series of tests with the FDR correction to properly implement simultaneous testing. For example, returning to the GLMM \eqref{ChenLi1} and \eqref{ChenLi2}, after computing the maximum likelihood estimates (MLE) of the parameters, likelihood ratio tests (LRTs) are also available to assess the effect of fixed effects on the presence of bacterial taxa ($H_0: \boldsymbol{\alpha}=\mathbf{0}$). The association of taxa with covariates can also be identified using the LRT for ($H_0: \boldsymbol{\beta}=\mathbf{0}$). This method can be used with any GLMM to correctly account for the effect of intra-subject correlation. The estimation and testing are available in \texttt{R} package \href{https://github.com/PennChopMicrobiomeProgram/ZIBR}{\texttt{ZIBR}}. 
\section{LonDA}\label{AppenLonDA}
Linear discriminant analysis (LDA) is also a widely used method in omics data analysis. The main aim of a classification problem is a prediction of the label of new samples based on the provided data. This problem becomes more complicated by adding the time effect in longitudinal data. Moreover, in an omics study and particularly in rare disease studies the number of subjects labeled `case' is severely outnumbered by `control' subjects, which tends to imbalanced samples. A methodology for the longitudinal discriminant analysis (LonDA) was introduced by Marshall and Bar\'{o}n \citep{Marshall2000}. Simply speaking, LonDA classifies the samples into predefined labels based on the observed LOD of each sample. In practice, each subject at each time-point is labeled, and the LonDA provides a model based on the observed training set of longitudinal data to predict the label of unlabeled subjects. In more detail, LonDA is a classification model that considers the change in the parameter set of distribution of each group across time. Let $G_j$, $j=1,\ldots,g$ denotes $g$ labels (populations) and $\mathbf{y}_i$ be the vector of observed values (data) for subject $i$ on arbitrary times $\mathbf{t}=(t_1,\ldots,t_{n_i})^T$ then the corresponding distribution of $\mathbf{y}_i$ would be $f_j(\mathbf{y}_i;\phi_j(\mathbf{t}))$ where the dimension of density parameters, $\phi_j(\mathbf{t})$, depends on the number of replicates (or dimension of $\mathbf{t}$) per each subject. For prior probabilities $\pi_1,\ldots,\pi_g$, the decision boundary \citep{ELSII} is 
\begin{eqnarray}\label{AppenESL:LDA}
	\log \pi_k +\log f_k(\mathbf{y}_i;\phi_k(\mathbf{t})) =\max_j\{	\log \pi_j +\log f_j(\mathbf{y}_i;\phi_j(\mathbf{t}))\}, ~~~~~~~j=1,\ldots,g. 
\end{eqnarray}
If the densities are Gaussian with the same covariance matrices, the decision boundaries are linear. Even in this case, the mean value of the $j^\text{th}$ population, $\boldsymbol{\mu}_j(\mathbf{t};\boldsymbol{\alpha}_j)$ say, can make a nonlinear relationship with the covariates where $\boldsymbol{\alpha}$ is the mean population specific parameter. The rest of the modeling procedure returns to the estimation of the parameters. For instance, if $\mathbf{V}_j(\mathbf{t},\boldsymbol{\alpha}_j;\boldsymbol{\theta}_j)$ be the variance matrix of vector $\mathbf{y}$ in the population $j$, where $\boldsymbol{\theta}_j$ is the variance specific parameter of population $j$, then regarding \eqref{AppenLMMVar} one can write
\begin{eqnarray*}
	\mathbf{V}_j(\mathbf{t},\boldsymbol{\alpha}_j;\boldsymbol{\theta}_j)= \mathbf{Z}(\textbf{t};\boldsymbol{\alpha}_j)\boldsymbol{\Sigma}_{\mathrm{rndEff}}(\boldsymbol{\theta}_j)\mathbf{Z}^T(\textbf{t};\boldsymbol{\alpha}_j)+\boldsymbol{\Sigma}_{\mathrm{Err}}(\boldsymbol{\theta}_j),
\end{eqnarray*}
where $\mathbf{Z}(\textbf{t};\boldsymbol{\alpha}_j)$ is the design matrix corresponding to the random effect in population $j$. Parameter vectors $\boldsymbol{\alpha}_j$ and $\boldsymbol{\theta}_j$ can be either identical with the mean and variance elements or link them to the covariates. Depends on the homogeneity of $\mathbf{Z}(\textbf{t};\boldsymbol{\alpha}_j)$ and the variance related parameters, $\boldsymbol{\theta}_j$, four estimation situations were expected \citep{Marshall2000}: 
\begin{enumerate}
	\item Homoscedastic Model: $\mathbf{Z}(\textbf{t};\boldsymbol{\alpha}_j)=\mathbf{Z}(\textbf{t})$ and $\boldsymbol{\theta}_j=\boldsymbol{\theta}$,
	\item Mean-heteroscedastic model: $\mathbf{Z}(\textbf{t};\boldsymbol{\alpha}_j)$ varies in $j$ but $\boldsymbol{\theta}_j=\boldsymbol{\theta}$,
	\item Variance-heteroscedastic model: $\mathbf{Z}(\textbf{t};\boldsymbol{\alpha}_j)=\mathbf{Z}(\textbf{t})$ but $\boldsymbol{\theta}_j$ varies in $j$, and
	\item Fully-heteroscedastic model: Both the $\mathbf{Z}(\textbf{t};\boldsymbol{\alpha}_j)$ and $\boldsymbol{\theta}_j$ vary in $j$. 
\end{enumerate}
The methodology was used in studying the subunit beta measurements of 174 pregnant women over 2 years in a private obstetrics clinic in Santiago, Chile \citep{Marshall2000}. The data were imbalanced since there was only one measurement for 30\% of subjects, two for 31\%, three for 33\%, and 6\% with more than three measurements. The subunit beta is used as a biomarker in this study to discriminate the samples with normal/abnormal delivery. Prior probabilities were the proportion of each group in the sample. The response variable was $\log$ subunit beta measurements and the response of the $i^\text{th}$ subject at time $t_{ik}$ when it belongs to $i^\text{th}$ group in the sample, $y_{i,t_k}^{(j)}$ for instance under the Mean-heteroscedastic model follows
\begin{eqnarray*}
	y_{i,t_k}^{(j)}=\frac{\alpha_{j1}+b_{ij}}{1+\alpha_{j2}e^{-\alpha_{j3}t_{ik}}}=\varepsilon_{it_{ik}j}, 
\end{eqnarray*}
where here, $i=1,\ldots,174$, $j=1,2$ and $t_ik$ is the $k^\text{th}$ time of measurement for subject $i$ where $k=1,\ldots,n_i$. In this study, $n_i=1,2,3,>3$ and suggests the notation $\boldsymbol{\alpha}_j=(\alpha_{1j},\alpha_{2j},\alpha_{3j})$. The remaining three models can be written in the same way and the fitted models chose the fully heteroscedastic model based on the deviance test which confirms the longitudinal effects in both the mean and variance of the $\log$ subunit beta measurements in two groups. Although there is no restriction of using the same method on the multivariate case, the bivariate case of this model was specifically discussed \citep{Marshall2009} with considering the missing completely at random structure applied on the Estradiol and $\beta$-HCG concentrations but using the 161 subjects and it is mentioned that they employed \texttt{NLMIXED} procedure from \texttt{SAS} to estimate the parameters. The same data same model under fully-heteroscedasticity assumption using the nonlinear hierarchical classification were studied in a fully Bayesian approach \citep{delaCruzMesia2007}. A clustering method was also presented \citep{Villarroeletal2009} based on a mixture decomposition scheme \citep{QinSelf2006} and then performed the same classification problem under both the balanced and imbalanced cases for multivariate longitudinal data. 

\section{Bayesian approach on LMM and GLMM}\label{AppendBayesian}		
Employing the Bayesian approach in longitudinal data analysis mostly happens through two main ideas: (1) A Bayesian statistical model (see Robert  \citep{Robert2007} Page 9 for the definition of the parametric version which considers a prior distribution on the parameters of a parametric statistical model) and (2) Bayesian learning as a machine learning paradigm by considering parameters as a latent variable \citep{Blei2017}. Using either approach, the main step is a posteriori computation and mostly arises when the sample size is not large enough to achieve the usual asymptotics of LMM and GLMM from the frequentist's viewpoint. The main step in the Bayesian framework involves incorporating the expert's opinion into the prior distribution in such a way that updating the posterior information via data does not result in unmanageable computational complexity. Here, we only restrict our review to the Bayesian practice of LMM and GLMM. In the sequel, we are listing the names of some of the approaches covered by computational tools with an application on LOD for each if exists. 

The problem of estimating the parameters of LMM in the Bayesian approach has been widely studied and performed in different software. We encourage the readers to consider the random effect selection in a hierarchical Bayesian framework \citep{ChenDunson2003}, which also gives the idea of Bayesian modeling of LMM. \texttt{brms} \citep{brms2021} introduced a fully Bayesian framework for the GLMMs with the possibility of zero-inflation and also provided theoretically defined identifiable models. A very practical version of this approach, only for GLMM with restricted distributions is conserved in \texttt{blme} \citep{Chungetal2013}. \texttt{JMbayes} \citep{Rizopoulos2016} is the Bayesian version of the joint modeling of LMM and survival models \texttt{JM} \citep{Rizopoulos2010}. \texttt{APML0} \citep{Xu2020} has been designed for longitudinal genome data and considers the phenotype as the response variable and the genome as a fixed effect. It provided a fast $\mathbb{L}_0$-penalized estimator for the genes effect and nominated the important genes as biomarkers. Bayesian model selection by considering the marginal and conditional criteria for count data is available in \texttt{BayesselectGLMM} function in \texttt{R} \citep{Ariyo2021} which provides a computationally adaptive method to compute the Bayes and pseudo-Bayes factors. The method was examined on clinical data; however, in genome-wide data, where usually each gene is modelled versus metadata, the same approach is applicable on the abundance table. The \texttt{Stan} stand-alone software \citep{Sorensen2016} also can be hired for Bayesian estimation of model \eqref{LMM} per each response variable. There is also a guideline to use \texttt{winBUGS} package in \texttt{R} for the general framework of Bayesian GLMM, which can be used for longitudinal data. \texttt{bmrarm} is designed to jointly model the longitudinal continuous and ordinal responses in the Bayesian domain \citep{Seedorff2022}. This can help mapping the longitudinal metadata onto baseline omics features. Some longitudinal omics studies although they did not employ a Bayesian analysis, they used the normalized counts generated by \texttt{DESeq2} \citep{Love2014}, which is essentially a Bayesian GLM estimation tool under negative binomial distribution. This approach can be found for single-cell data \citep{Tsangetal2017} and in previously discussed transcriptome data \citep{Sacramentoetal2020}.

\section{Survival, dropout and sparsity}\label{AppenSurvival}

In omics studies, particularly for severe diseases, subjects may leave before the study ends due to treatment changes, withdrawal, recovery, or death. These scenarios require specific modeling, often incorporating the time of exit. When the study focuses on this time (e.g., recovery/mortality), survival models are used. A key challenge is predicting disease outcomes based on LOD, as survival data is often sparse, with each subject measured only once. However, some studies integrate survival information with longitudinal measurements to model disease progression and predict survival. Notable examples include survival prediction using CD4 counts in AIDS cohorts \citep{Tsiatis1995,WangTaylor2001} and PSA levels in prostate cancer \citep{Yuetal2004}. While survival datasets record time-to-event, single-time measurements are not considered longitudinal. This review focuses on survival studies with multiple time-point measurements.

Most longitudinal omics studies employ Cox regression for survival analysis, focusing on time-to-event data \citep{Sacramentoetal2020}. While many use Cox models independently of omics data, some integrate joint modeling to assess the impact of longitudinal omics and baseline metadata, or baseline omics and longitudinal clinical data, on hazard rates.

\subsection{Joint Linear Mixed and Cox Regression Models}

In longitudinal treatment studies, joint models assess repeated measures while linking covariate effects (e.g., treatments) to time-to-event outcomes, such as biomarker level-crossing or dropout. A classic example is Zidovudine (ZDV) treatment for HIV, where CD4 counts serve as surrogate markers for survival \citep{Tsiatis1995}. A double-blind placebo-controlled trial with 281 advanced HIV patients (144 receiving $250mg$ ZDV every four hours and the rest receiving placebo) measured CD4 levels periodically. The survival time $T^*$ was modeled using CD4 levels ($y_{ij}$) and treatment within an LMM \eqref{LMM}. With right-censoring at $T = \min\{T^*, C\}$, Cox regression estimated the hazard function $\lambda(t|\mathbf{y})=\lambda_0(t)g(\mathbf{y},\boldsymbol{\gamma})$. Nonrandom missingness introduced bias, as lower CD4 levels increased dropout probability. The random effect (time trend) was estimated via empirical Bayes \citep{LairdWare} adjusted for missing. The \texttt{brms} package in \texttt{R} estimates model parameters without missing data adjustments \citep{Tsiatis1995}. Later, an EM algorithm was introduced for joint LMM-Cox regression estimation, yielding comparable results \citep{WulfsohnTsiatis1997}. The extended change-point model is discussed in the next section.

The \texttt{JSM} package \citep{Xu2020JSM} jointly models longitudinal data using LMM and transformation models for survival analysis. The random effect function is estimated nonparametrically, and the baseline hazard function is modeled semiparametrically. Biomarkers serve as response variables, incorporating baseline data as fixed effects.

\subsection{Integrated Ornstein-Uhlenbeck Random Effect}

The two-stage joint model \citep{Tsiatis1995} was extended to include time-varying coefficients in CD4 studies. Data from 115 seroconverters estimated infection time as the median between the last negative and first positive HIV test, excluding early CD4 measurements due to abrupt changes. Disease progression or last visit was modeled as:
\begin{eqnarray*}
	\left\{
	\begin{array}{lcl}
		y_{ij}&=&\psi_i(t_{ij})+\varepsilon_{ij},\\
		\psi_i(t)&=&a_i+ bt+ \mathbf{X}_i\boldsymbol{\beta}+ W_i(t),
	\end{array}\right.
\end{eqnarray*}
where $a_i$ is the random intercept, and $W_i(t)$ follows an Ornstein-Uhlenbeck process. The hazard function 
$\lambda(t|\phi_i(t),\mathbf{X}^*)=\lambda_0\exp(\gamma\psi_i(t)+\mathbf{X}^*\boldsymbol{\beta}^*)$ was estimated via Bayesian approach and MCMC \citep{WangTaylor2001}. The dataset was highly imbalanced (1–32 visits per subject). Adjusting for $\psi_i$, baseline $\sqrt[4]{\mathrm{CD4}}$, and age had minimal effects in the Cox model. Results confirmed declining CD4 counts negatively impacted survival ($b, \gamma < 0$), with reduced MSE compared to random effect models.

\subsection{Change-Point Detection Model}

Plasma HIV RNA data from the AIDS Clinical Trials Group 398 study \citep{Hammer2002} assessed virologic failure after 24 weeks in 481 patients across 12 time points over 72 weeks, comparing single and dual protease inhibitor (PI) regimens. Dropout and missing data led to an imbalanced design. Viral load trajectories were modeled as:
\begin{eqnarray*}
	y_{ij}=\mathbf{X}_i\boldsymbol{\beta}+ b_{i0}+b_{i1}t_{ij}+ \sum_{l=1}^{L}\alpha_{il}(t_{ij}-\tau_l)_++\varepsilon_{ij}\coloneqq \psi_{i}(t_{ij})+\varepsilon_{ij},
\end{eqnarray*}
where the random effect design matrix is linear in time, with piecewise behavior at $L$ change-points \citep{Ghosh2011}. The error term follows $\varepsilon_{ij} \sim N(0,\sigma^2/\eta_i)$, allowing subject-specific variability.

Dropout was modeled using a Cox proportional hazards model $\lambda(t|\psi_i)= \lambda_0(t) \exp(\gamma\psi(t))$, where larger $\gamma$ then a higher dropout rate. Parameters were estimated using a Bayesian MCMC approach, assuming missing-at-random data. Covariates included treatment (four arms: three dual-PI, one placebo-PI) and non-nucleoside reverse transcriptase inhibitor (NNRTI). Results captured a significant NNRTI effect and a positive association between dropout and RNA viral load.

\section{Supplementary figures}
\begin{figure*}
\centering
\includegraphics[width=.98\textwidth]{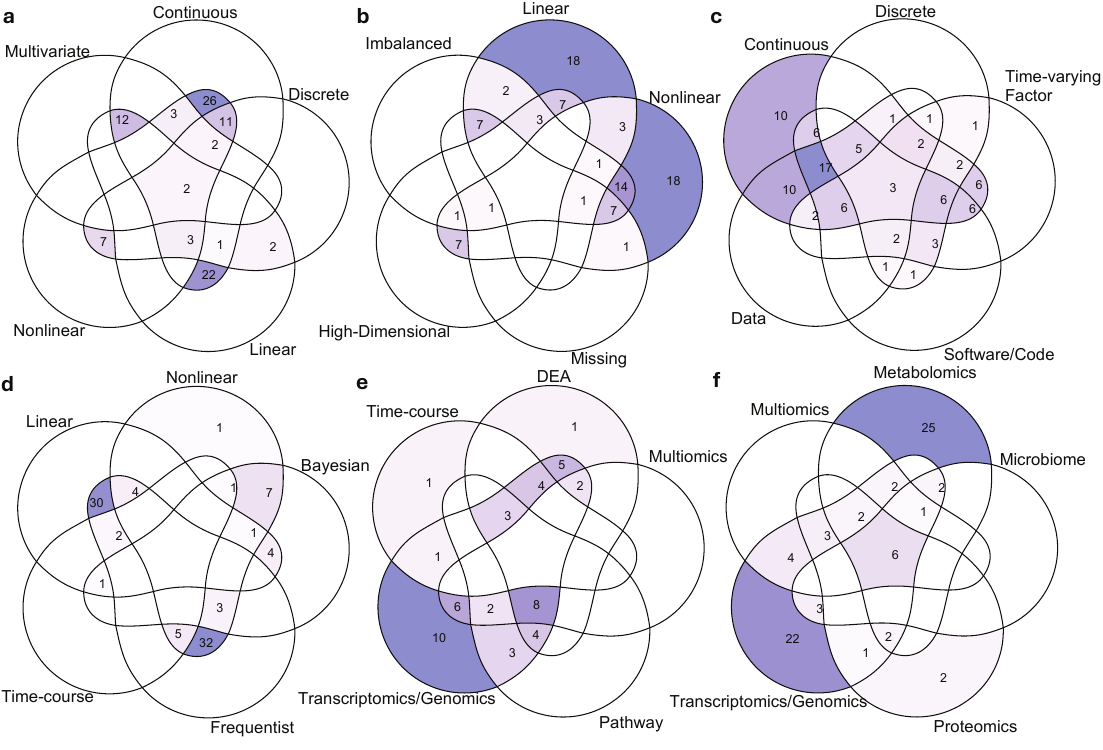}
\caption{The number of each study with LOD we have considered from different points of view. \textbf{a}, shows how studies cover the analyses by dealing with the abundances as discrete or continuous variables. It also differentiates between the methodologies that use multivariate approaches rather than the ones that use multiple univariate methods. \textbf{b}, demonstrates the same idea from the sample design viewpoint and categorizes the references if they have considered the effect of imbalanced sample designs and missing values. \textbf{c}, shows the availability of the computational tools or codes as well as the publicly available data across the variable types (count/relative abundance) by categorizing the fixed effects as time-varying or observed only at the baseline. \textbf{d}, shows the frequencies of approaches used by the studies from Bayesian or frequentist viewpoints. It also shows the frequency of time-course studies. \textbf{e}, categorizes the studies from the type of biological approach. \textbf{f}, demonstrates the distribution of studies across the omics types by considering the multiomics studies.}\label{figsummaries}
\end{figure*}

\begin{figure*}
\centering
\includegraphics[width=1\linewidth]{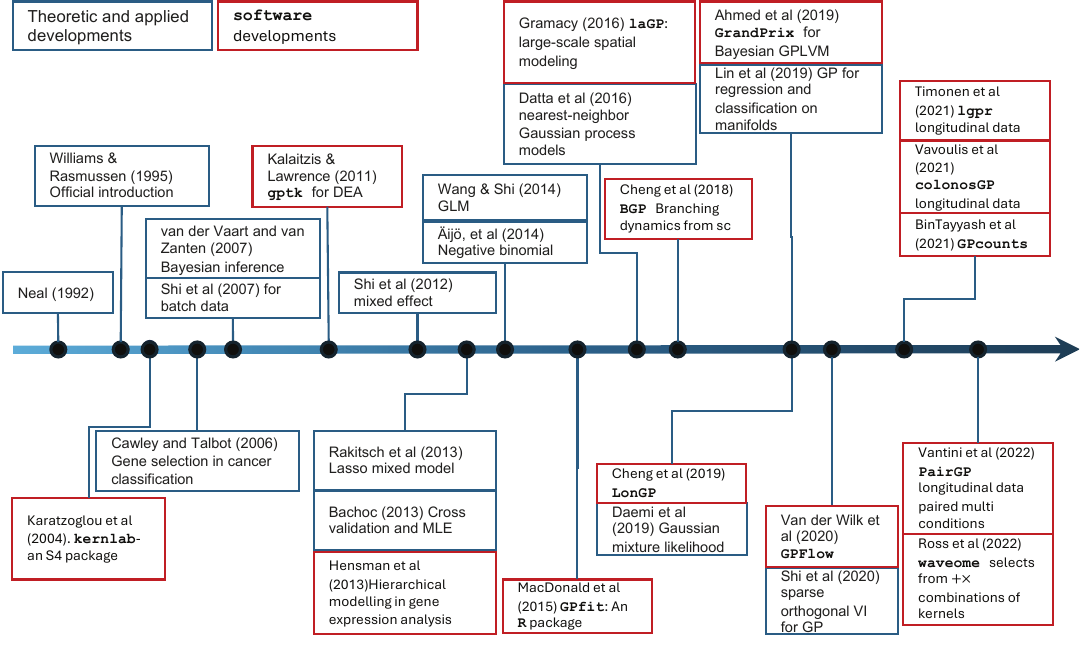}
\caption{The evolution of GP approach across the time \citep{Neal1992,WilliamsRasmussen1995,Karatzoglouetal2004,CawleyTalbot2006,VanderVaartVanZanten2007,Shietal2007,Kalaitzisetal2011,Shietal2012,Rakitschetal2012,Bachoc2013,Hensmanetal2013,WangShi2014,Aijoetal2014,MacDonaldetal2015,Dattaetal2016,Gramacy2016,Boukouvalasetal2018,Ahmedetal2019,Linetal2019,Daemietal2019,Chengetal2019,vanderWilketal2020,Shietal2020,BinTayyashetal2021,Timonenetal2021,Vavoulisetal2021,Vantinietal2022,Rossetal2022}.}
\label{fig:GPmodels}
\end{figure*}

\end{document}